\documentclass[noshowpacs,prl,twocolumn,preprintnumbers,superscriptaddress,amsmath,amssymb,floatfix]{revtex4}
\usepackage[caption=false]{subfig}
\usepackage{graphicx}
\usepackage{placeins}
\usepackage{tikz}
\usepackage{mathtools}
\usepackage{amsthm}
\usepackage{romannum}
\usepackage{color,soul}
\usepackage{xkeyval,xcolor}
\usepackage{chngcntr}

\newcommand{\nn}{\langle \mathcal{N} \rangle}

\begin{document}
	
	\title{Lifetime of a greedy forager with long-range smell}
	\author{Hillel Sanhedrai}
	\affiliation{Department of Physics, Bar-Ilan University, Ramat Gan, Israel}
	\author{Yafit Maayan}
	\affiliation{Department of Mathematics, Jerusalem College of Technology (JCT), Jerusalem, Israel}
	\author{Louis Shekhtman}
	\affiliation{Department of Physics, Bar-Ilan University, Ramat Gan, Israel}
	\date{\today}
	
	\begin{abstract}
		We study a greedy forager who consumes food throughout a region. If the forager does not eat any food for $S$ time steps it dies. We assume that the forager moves preferentially in the direction of greatest smell of food.  Each food item in a given direction contributes towards the total smell of food in that direction, however the smell of any individual food item decays with its distance from the forager. We assume a power-law decay of the smell with the distance of the food from the forager and vary the exponent $\alpha$ governing this decay. We find, both analytically and through simulations, that for a forager living in one dimension, there is a critical value of $\alpha$, namely $\alpha_c$, where for $\alpha<\alpha_c$ the forager will die in finite time, however for $\alpha>\alpha_c$ the forager has a nonzero probability to live infinite time.  We calculate analytically, the critical value, $\alpha_c$, separating these two behaviors and find that $\alpha_c$ depends on $S$ as $\alpha_c=1 + 1/\lceil S/2 \rceil$. We determine analytically that at $\alpha=\alpha_c$ the system has an essential singularity. We also study, using simulations, a forager with long-range decaying smell in two dimensions (2D) and find that for this case the forager always dies within finite time. However, in 2D we observe indications of an optimal $\alpha$ for which the forager has the longest lifetime.
		
	\end{abstract}

\maketitle

\pagenumbering{arabic}

\section{Introduction}

Animals seeking food or resources spread out over a region often must move throughout the region in order to obtain the desired resources. The question of whether such foraging can be performed in an optimal manner to maximize the animal's lifetime or likelihood of finding food has received significant attention \cite{stephens1986foraging,pyke1984optimal}. Many earlier studies argue that to be optimal such searching should be done stochastically \cite{oaten1977optimal,green1984stopping} and that random walks or L\'evy flights can be used to model this behavior \cite{viswanathan1996levy,viswanathan1999optimizing,benichou2005optimal,lomholt2008levy}. 

Recent work has suggested a new model where  a forager carries out a random walk, yet the food is explicitly consumed until the forager starves to death \cite{benichou2014depletion,benichou2016role}. In this model, the forager begins at some point on a lattice with each site containing food. The forager then moves and eats the food at the discovered site, leaving no remaining food there. It continues to move throughout the region either returning to sites without food or eating food at new discovered sites. If the forager goes $S$ steps without eating, it starves to death. Notably, this process leads to inherent desertification \cite{reynolds2007global,weissmann2014stochastic}, as the forager eventually creates a desert of visited sites among which it moves until starvation . Later work has expanded this to cases where the food renews after some time \cite{chupeau2016universality}, where the forager eats only if it is near starvation \cite{rager2018advantage,benichou2018optimally}, and where the forager walks preferentially in the direction of a \emph{nearby} site with food (greed) \cite{bhat2017does,bhat2017starvation}. 

\begin{figure}[ht]
	\includegraphics[width=0.49\textwidth]{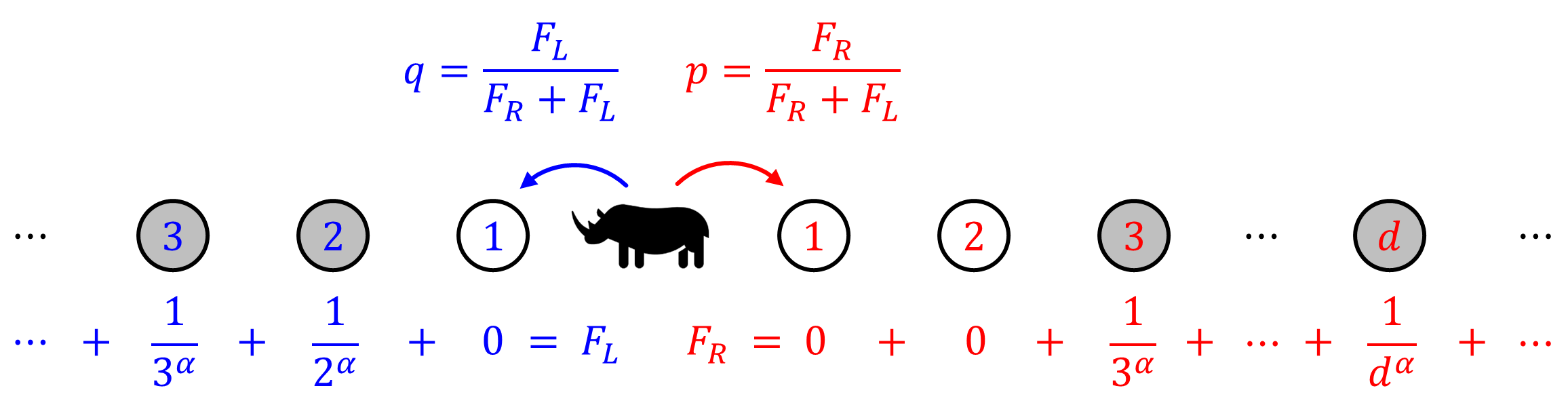}	
	\caption[short text]{Model demonstration in 1D. We show here how the likelihood of the forager to walk in a given direction depends on the amount of food in that direction.  In one dimension there are only two different directions in which the total smell, $F_{R(L)}$ must be calculated and $d$ is simply given by the distance of the forager from the food. The probability to walk right (left) is then given by $p(q)$ as defined in the figure. }
	\label{fig:model}
\end{figure}

Here we build on these recent models of a starving forager by considering a forager with an explicit sense of \emph{smell}. In this sense, we assume that in each direction  the forager can move, there is some total  smell represented by the sum of contributions from each site with food. We assume that the contribution of an individual site with food to the overall smell in a given direction decays with its distance $d$ from the forager, as $d^{-\alpha}$ where $\alpha$ controls the rate of the decay. We then assume that the forager walks probabilistically in each direction proportional to the total smell in that direction. We note that in the limit of $\alpha\to\infty$ only the nearest site affects the sum and thus the results of \cite{bhat2017does,bhat2017starvation} for a greedy forager are recovered. However, in contrast to their study, our model incorporates the influence of food at distances greater than 1, though with an impact that decays based on the distance, see Fig.~\ref{fig:model}.

\section{One Dimension}
\begin{figure*}[htbp]
	\centering

	\subfloat{%
		\includegraphics[width=0.32\linewidth]{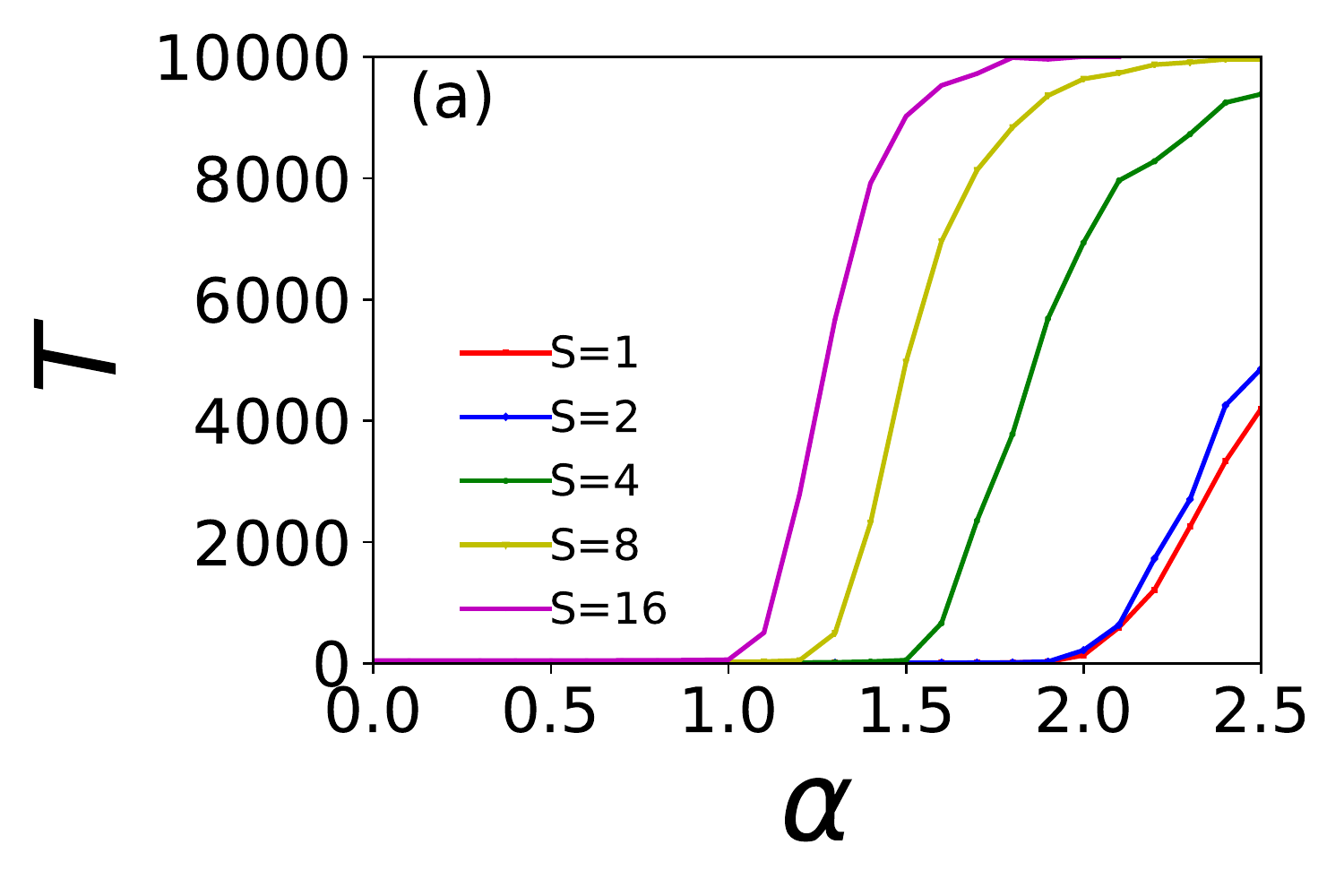}  		\label{fig:1d-finite}
	}\hfill
	\subfloat{%
		\includegraphics[width=0.32\linewidth]{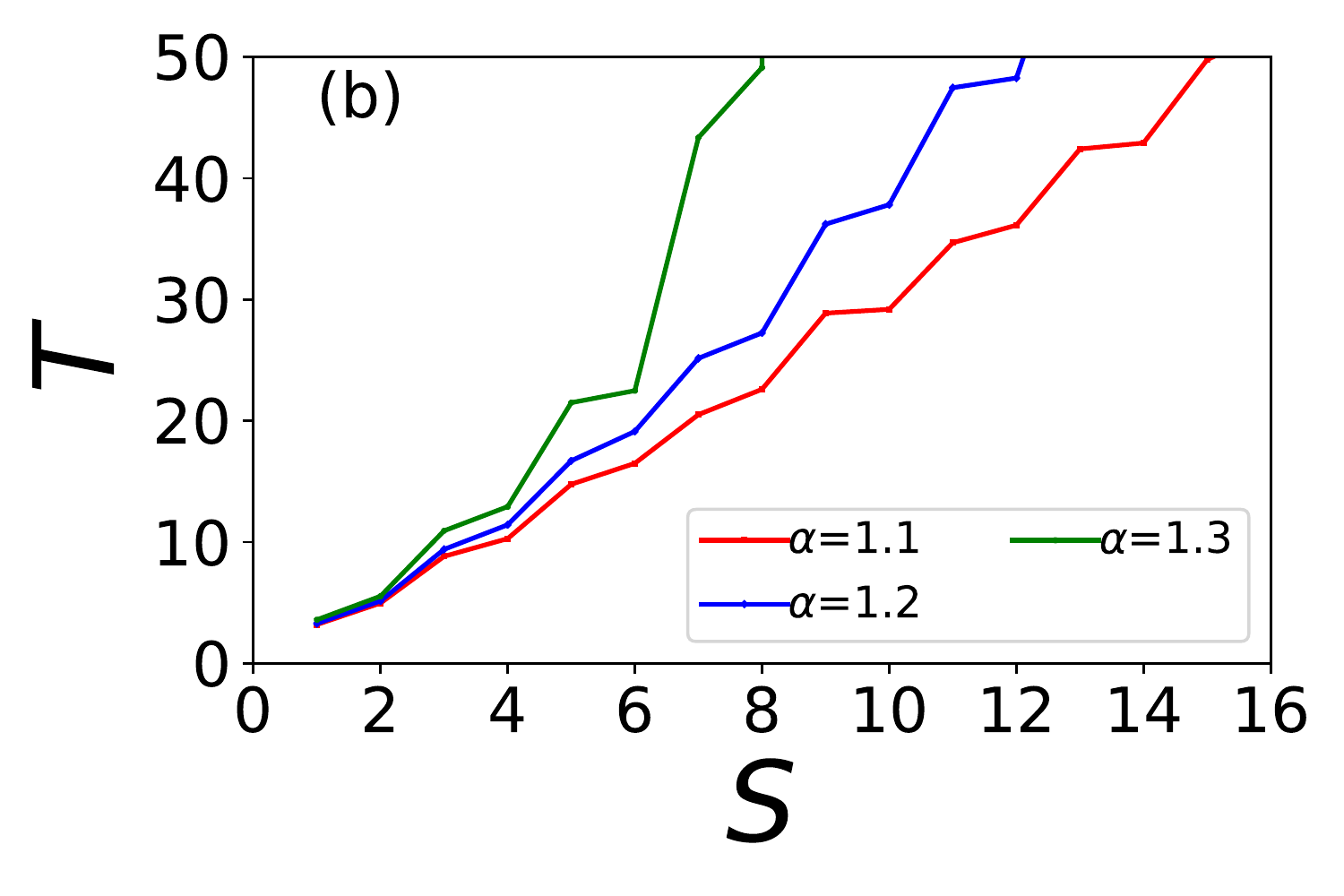}  		\label{fig:1d-T-v-S}	
		
	} \hfill
	\subfloat{
		\includegraphics[width=0.32\linewidth]{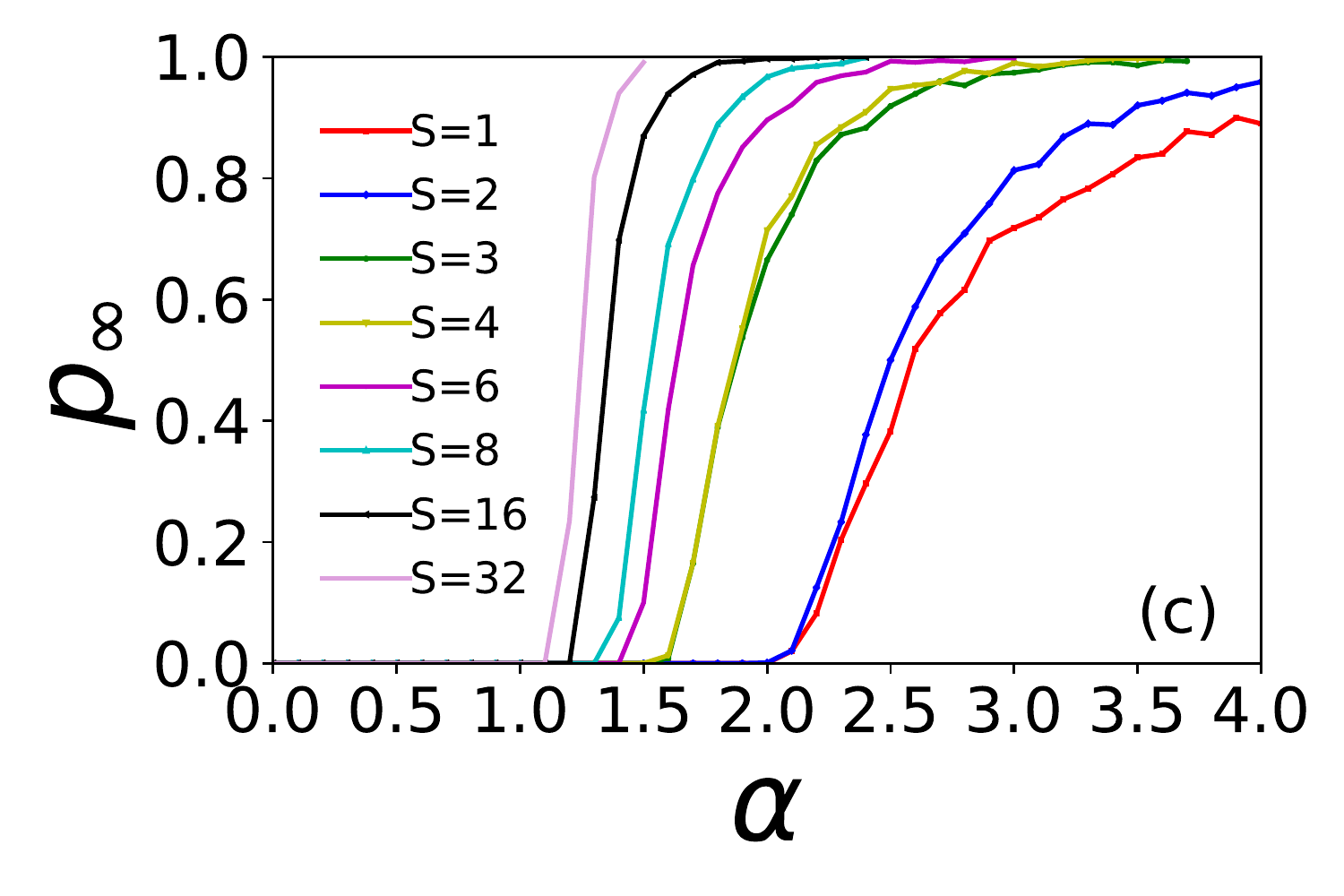} 		\label{fig:1d-pinf}
		
	}
	
	\caption{Plots for a greedy forager with long-range smell in one dimension, Eq.~\eqref{eq:smell-def}. In \textbf{(a)} we show simulation results for the mean lifetime of the forager, $T$, as a function of $\alpha$ for several $S$ values for a one dimensional line of $l=10^4$ sites with periodic boundary conditions. We see that as $\alpha$ increases the forager reaches a point where it eats all of the food. All runs are averaged over $1000$ realizations. \textbf{(b)} Here we plot $T$, the forager lifetime, as a function of $S$ for several fixed $\alpha$. We choose $\alpha$ values just above 1 in order to obtain finite values of $T$ for values of $S$ before the forager reaches a point where $T \to \infty$. We observe an even-odd alternating pattern. 
	\textbf{(c)} We plot the probability of the forager to live for an infinite time, $p_\infty$. In practice, we carry out $1000$ realizations and find the fraction of those that approach to $T\to\infty$. Since we must stop the simulations at some finite point, we set $T=10^7$ as the cutoff point.  }
	\label{fig:forager-prob}
\end{figure*}
Like the earlier studies of \cite{benichou2014depletion,bhat2017does}, we begin with the case of one dimension, see Fig.~\ref{fig:model}. This case is more tractable to analytic solutions and provides intuition for the more ecologically relevant case of two dimensions.

We assume that the forager begins at the center of a one dimensional lattice of length $l$ with periodic boundary conditions. 
We can start by defining the total smell of food in a given direction as
\begin{equation}
F=\sum_{d=1}^{l/2}\frac{\delta_d}{d^\alpha}
\label{eq:smell-def}
\end{equation}
where $d$ is the distance of the forager to the site being considered, $\delta_d$ is $1$ if the site at distance $d$ contains food and $0$ otherwise, and $\alpha$ is the parameter controlling the decay of the smell with distance. Ideally, we are interested in the limit of $l \to \infty$, but we will start by considering a finite $l$ and then take the limit $l\to \infty$.

For one dimension there are two directions in which the forager may move and thus two directions of total smell. We denote these as $F_R$ being the smell to the right and $F_L$ being the smell to the left. The forager then selects in which direction to move with probability proportional to the total smell according to
\begin{equation}
\begin{cases}
p = \frac{F_{R}}{F_R+F_L},\\
q = \frac{F_{L}}{F_R+F_L}
\end{cases}
\label{eq:prob-walk-1d}
\end{equation}
where $p,q$ refer to the probability to walk in the right or left direction respectively (see Fig.~\ref{fig:model}). Thus, at each step the forager has a bias to walk in the direction with more total food at a closer distance.

We present simulation results for the system described above in Fig.~\ref{fig:1d-finite} for different values of $S$ and measure the mean forager lifetime $T$ as a function of $\alpha$. We observe that up until around $\alpha\approx 1$ the forager remains at a fairly small number of steps (though not zero, rather simply below $\approx 100$), see Fig.~\ref{fig:1d-T-v-S}. We understand this to be the result of the fact that for these small values of $\alpha$ the forager is highly considering food from very far away leading to a large total sum of food in both directions and causing the forager to essentially carry out an unbiased random walk as in \cite{benichou2014depletion}.

In contrast, we see that above a certain $\alpha$, $\alpha_c$, which decreases with $S$, the forager consumes all (or a large fraction of) food in the system, i.e. $T\approx 10,000=l$. This is due to the fact that for sufficiently large $\alpha$, once the forager takes one step in either direction, it will then mostly consider its immediate neighbors with increasing probability with time. The neighboring site of the forager's previous position will be empty, while the other neighboring site will be filled, thus the forager will move towards the filled site. This will happen nearly every time step and the forager will simply continue moving in the same direction and eating food at new sites. Even if the forager does happen to take a single step back, it will likely quickly return to its path.  In the limiting case of $\alpha\to\infty$, our results approach those of \cite{bhat2017does} with perfect greed since this limit considers only the nearest site.

Using the intuition gained above, we now consider the case of $l\to\infty$. For this case, we note that  a forager who just ate, will inherently be at the edge of a semi-infinite line of food in one direction. Therefore the overall smell from the direction of this line of food is
\begin{equation}
	F=\sum_{d=1}^\infty\frac{1}{d^\alpha}=\zeta(\alpha),
	\label{eq:smell-inf-1d}
\end{equation}
where $\zeta(\alpha)$ is the Riemann-Zeta function. 

Relating the total smell in a direction to the Riemann-Zeta Function (RZF) explains why $\alpha\leq 1$ recovers the results of \cite{benichou2014depletion}. This is since the the RZF diverges for $\alpha\leq 1$ giving infinite smell in both directions and thus equal probability for moving in either direction. 

This mapping to the $\zeta$ function also allows us to actually simulate a system of infinite size rather than merely increasing the value of $l$. This is because rather than looping over an infinite amount of food, we can subtract from $\zeta(\alpha)$ those locations which do not have food since the forager has already visited them. These locations are between the maximal and minimal sites that the forager has visited. Essentially, the smell in a given direction can be related to the difference between the forager's current location, $x_0$, and the maximal (minimal) location reached in that direction $x_\text{max (min)}$, such that
\begin{equation}
	F_i=\zeta(\alpha)-\sum_{d=1}^{x_\text{max (min)}-x_0}\frac{1}{d^\alpha}
	\label{eq:calc-f-1d-inf}
\end{equation}
where if $x_\text{max (min)}=x_0$, the sum is defined as $0$ and $F=\zeta(\alpha)$. After calculating the values of $F$ in each direction we again use Eq.~\eqref{eq:prob-walk-1d} to calculate the probability to move in each direction.

Next, we examine, in Fig.~\ref{fig:1d-T-v-S} the forager lifetime, $T$, as a function of $S$ for values of $\alpha$ slightly above 1, where the forager lifetime is finite. In Fig.~\ref{fig:1d-T-v-S}, we observe even-odd alternating steps in the forager lifetime. We recognize this as being due to the fact that if a forager has eaten for several steps in one direction and then takes one step away from the food, it will also not eat on the next step since both locations next to it will be empty. Thus, the forager requires at least 3 steps to return to new food (one to step away from the food, one to step back to its original location, and a third step to reach the new food). Similarly, if the forager takes two steps in the direction away from the food, then it will need 5 steps to return to new food and so on. Therefore only for odd $S$ does the forager lifetime increase significantly. 

Also, in analogy to the finite $l$ case where the forager consumes all food, for the infinite $l$ case we can expect that for sufficiently large $\alpha$ the forager will survive for infinite time. To calculate via simulations the likelihood of this occurring, a cutoff point at which we stop the simulations, must be chosen. In Fig.~\ref{fig:1d-pinf}, we plot the likelihood, $p_\infty$ that the forager will have a lifetime above the cutoff versus $\alpha$ for different values of $S$. We observe that there again appears some critical value of $\alpha$, $\alpha_c$, for which the forager lives an infinite lifetime. We note that it is somewhat surprising that $\alpha_c$ changes with $S$ as typically critical exponents are independent of the values of microscopic parameters.

\subsection{Theoretical Calculation of $\alpha_c$ }

\begin{figure*}[htp]
	\centering	
	\subfloat{
		\includegraphics[width=0.43\textwidth]{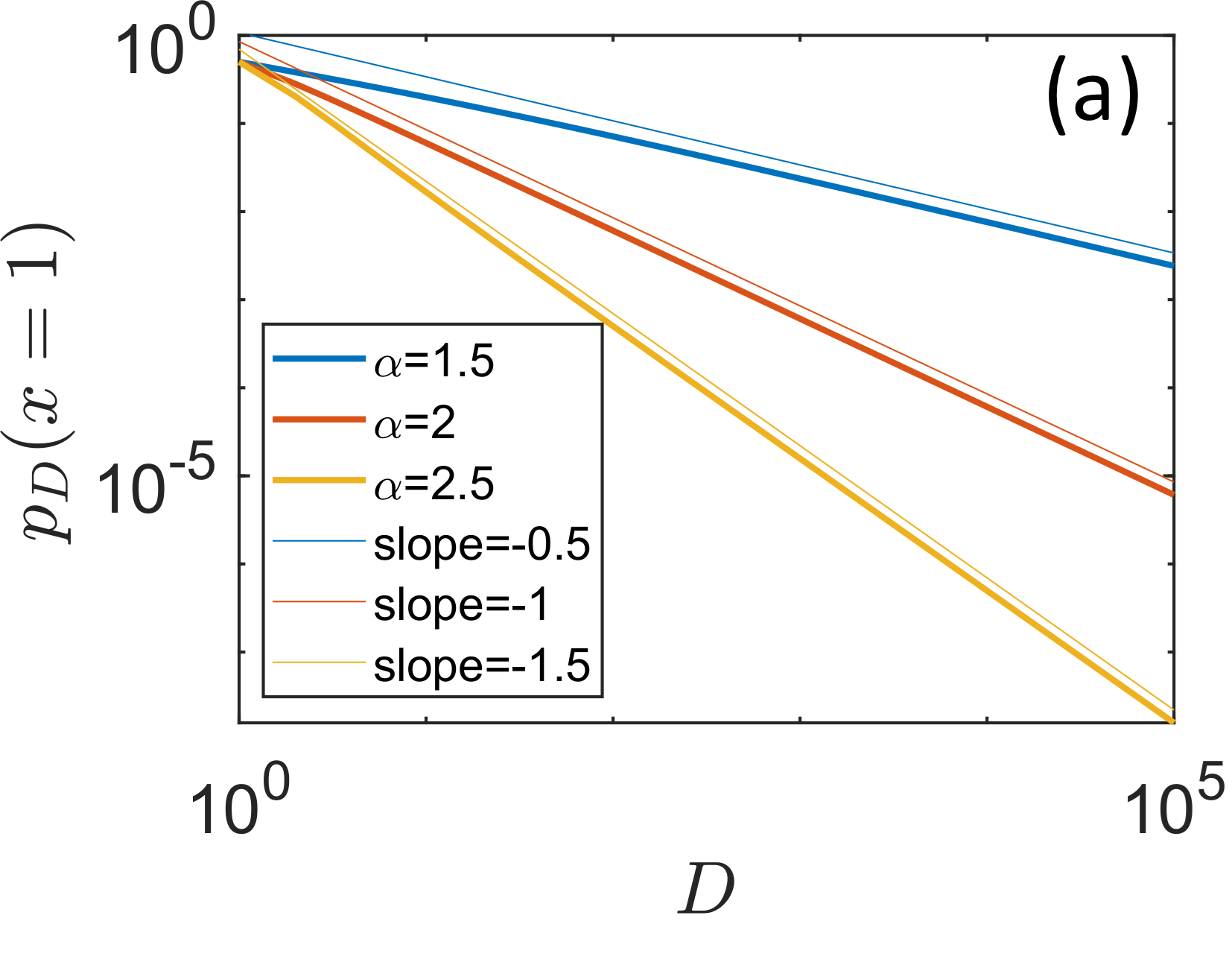}	
		\label{fig: p vs D x1}
	} \hfill
	\subfloat{
		\includegraphics[width=0.54\textwidth,trim={0 0.5cm 0 0  0},clip]{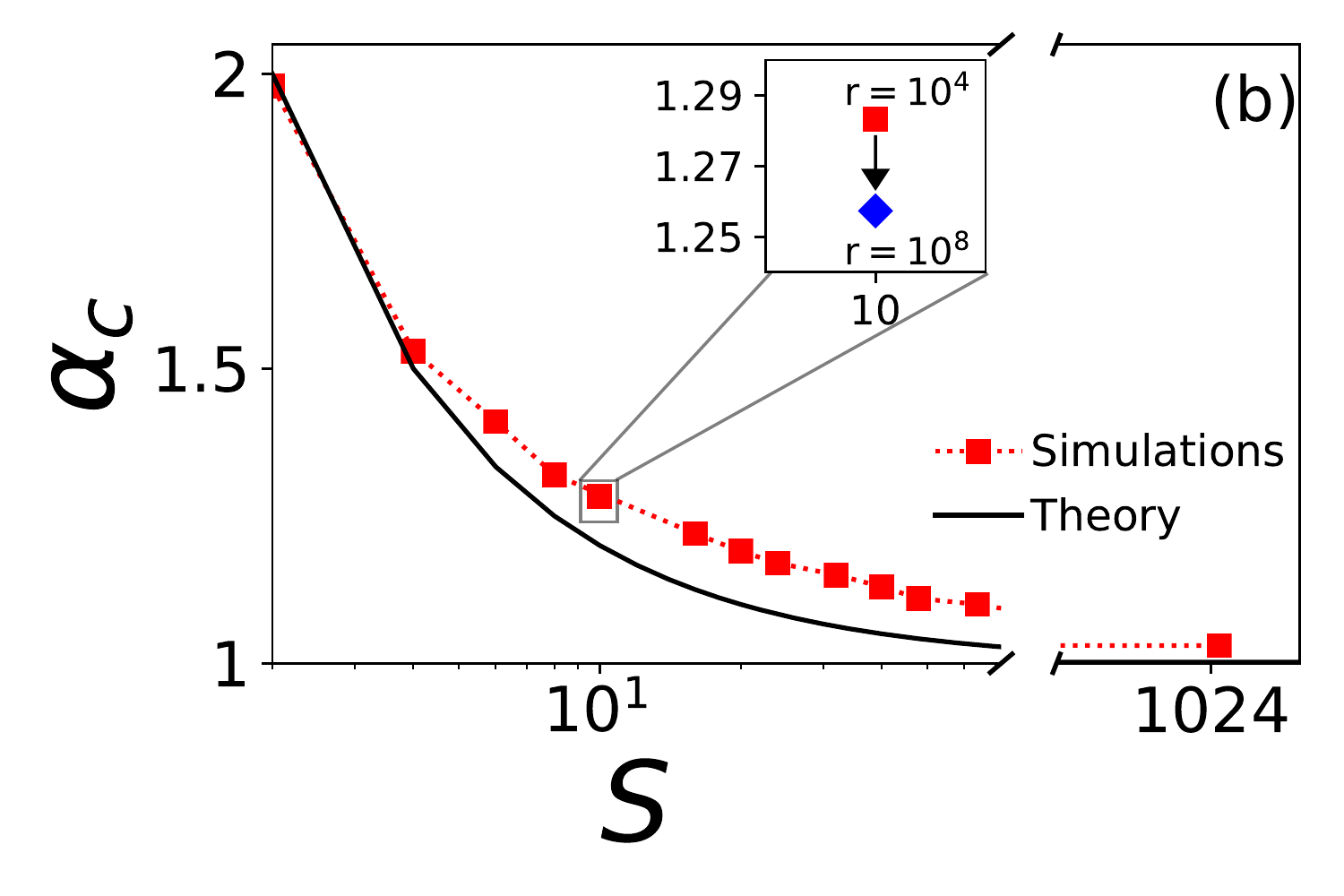} 
		\label{fig:1d-alphac}
	}
	\caption[short text]{
		\textbf{(a)} Here we show the likelihood of a forager with an infinite line of food  in one direction and $D$ empty food sites in the other direction (after which all sites have food), to choose to walk in the direction of the infinite line of food. We show this on a log-log plot for $x=1$. The dotted lines represent slopes with the values given in the legend. The scaling properties confirm the analytic result of Eq.~\eqref{eq:pL}. \textbf{(b)} We compare theory and simulations of $\alpha_c$ as a function of $S$. The simulations suggest values of $\alpha_c$ slightly larger than those calculated by the theory, yet this is likely due to the finite number of realizations, $r=10^4$, for each point and the likelihood of a single realization to reach $T\to\infty$ may be less than this value near $\alpha_c$. In the inset, we show that increasing the number of realizations (to $10^8$) leads to a decrease in the calculated value of $\alpha_c$, closer to the analytic result of Eq.~\eqref{eq: alpha_c vs s}. 
	}
	\label{fig: p vs L}
\end{figure*}

Having observed that the forager could live for $T\to\infty$ steps, we aim to calculate the lowest value of $\alpha$, $\alpha_c$, that this occurs. We define $\alpha_c$ such that for $\alpha>\alpha_c$ the probability to live forever ($T\to\infty$) is $p_{\infty}>0$ and the average number of distinct sites visited (food consumed), is $\nn\to\infty$, , while for $\alpha<\alpha_c$, $p_\infty=0$. If a forager lives forever, it will do so by almost always moving in the same direction, creating a desert of length $D$ between itself and food on the other side of the desert. After a large number of steps where the forager remains alive it will reach a point where $D$ is very large. 
The behavior in this limit will determine if the forager can live forever. 
%

To derive $p_\infty$ we will calculate the likelihood $\phi$ of the forager to survive until its next meal given that it just ate. We note that $\phi$ depends on $D$ and that each time the forager eats a meal, the size of the desert will increase by 1, $D\to D+1$.
We can now recognize that $p_{\infty}$, the probability of the forager to live forever, is
\begin{equation}
p_{\infty} = \prod_{D=1}^{\infty}\phi_D,
\label{eq: pinf}
\end{equation}
where $\phi_D$ represents the value of $\phi$ for a given $D$.
\noindent
Next, we calculate the values of $\phi_D$.
Without loss of generality, we will assume that  the forager has so far moved to the left and thus after its most recent meal, the desert of size $D$ is to its right. Between meals, the forager will wander and its distance from the next meal $x$ will vary i.e., it will move away from and back towards the edge of the desert.   
To calculate $\phi_D$, we must first determine the likelihood of the forager to move either towards or away from the edge of the desert. As in Fig.~\ref{fig:model} we denote as $p_D(x)$ the probability to move right (further into the desert) given that the desert is of size $D$ and that the forager is at distance $x$ from the desert's edge. Likewise, $q_D(x)$ is the likelihood to move left or towards the edge of the desert of size $D$. 

In the limit of large $D$, we can approximate  $p_D(x) = F_R/(F_R+F_L) \approx F_R/F_L $ since the forager is very far from the food at the opposite end of the desert (right side) and $F_R\ll F_L$. For a given value of $x$, $F_L$ will be the same  for any value of $D$. Thus, all that remains to find $p_D(x)$, is to determine $F_R = \sum_{n=(D-x)}^{\infty}\frac{1}{n^\alpha}$. This can be approximated by the integral
\begin{equation}
F_R\approx\int_{D-x}^{\infty}n^{-\alpha} \mathrm{d}n = \frac{(D-x)^{1-\alpha}}{\alpha-1}.\nonumber
\end{equation}
We note that to leading order this implies the scaling relationship $F_R \sim D^{1-\alpha}$ and so 
\begin{equation}
p_D(x) \sim F_R/F_L \sim D^{1-\alpha}.
\label{eq:pL}
\end{equation}
This scaling relationship is important for the remainder of the derivation and so we verify it for $x=1$ in Fig. \ref{fig: p vs L}a. 

\noindent
After finding the scaling of $p_D(x)$, Eq.~\eqref{eq:pL}, we now return to $\phi_D$ and its complement $1-\phi_D$.
We will consider `paths of starvation' i.e., those paths along which the forager will fail to consume its next meal and die. These paths must end at $x>0$ and thus they include only paths with equal or more steps to the right than to the left. This means that at least $k = \lceil S/2 \rceil $ steps must be to the right where $\lceil \cdot \rceil$ is the nearest integer above the number. Since $p_D(x)\ll 1$, we will only consider the leading term which consists of paths with the minimum number of steps to the right. Using Eq. (\ref{eq:pL}), we can write 
\begin{equation}
1-\phi_D \sim p_D^k q_D^{S-k} \sim D^{(1-\alpha)k}.
\label{eq: FL}
\end{equation}

\noindent
Knowing the scaling of $\phi_D$, we can now evaluate $p_{\infty}$ using Eq. (\ref{eq: pinf}).
We estimate $p_\infty$ as
\begin{align*}
\ln p_{\infty} &= \sum_{D=1}^{\infty}\ln \phi_D \\
&= \sum_{D=1}^{D_0-1} \ln \phi_D + \sum_{D=D_0}^{\infty} \ln (1-AD^{(1-\alpha)k}) 
\end{align*}
where $A$ is a constant prefactor and for $D\geq D_0$  we have $AD^{(1-\alpha)k}\ll 1.$ Thus,
\begin{equation}
\ln p_{\infty}\approx 
-B-A\sum_{D=D_0}^{\infty} D^{(1-\alpha)k} \label{eq:ln-pinf}\\
\end{equation}
with $B=\sum_{D=1}^{D_0-1} \ln \phi_D$  being another constant factor. The result in Eq.~\eqref{eq:ln-pinf} depends on if the sum diverges, which will depend on the value of $(1-\alpha)k$ such that
\begin{equation}
\ln p_\infty= \nonumber \begin{cases}
-\infty, \hspace{0.75cm} (\alpha-1)k \leq 1\\
-C(\alpha), \quad (\alpha-1)k > 1,
\end{cases}
\end{equation}
where $C(\alpha)$ is some finite value resulting from the infinite sum. 
Finally, we can obtain
\begin{equation}
p_{\infty} =
\begin{cases}
0, \hspace{1.18cm} \alpha \leq 1+\frac{1}{k}\\
e^{-C(\alpha)}, \quad \alpha > 1+\frac{1}{k}
\end{cases}
\label{eq: pinf critic}
\end{equation}
Thus, we find  $p_\infty>0$ for
\begin{equation}
\alpha>\alpha_c= 1 + \frac{1}{\lceil S/2 \rceil}.
\label{eq: alpha_c vs s}
\end{equation}
We show our analytic result compared to simulations in Fig.~\ref{fig:1d-alphac}. We note that in the limiting case $S=1$ then $\alpha_c=2$, and for $S$  large $\alpha_c \to 1$. 

\begin{figure}
	\centering
		\includegraphics[width=1\linewidth]{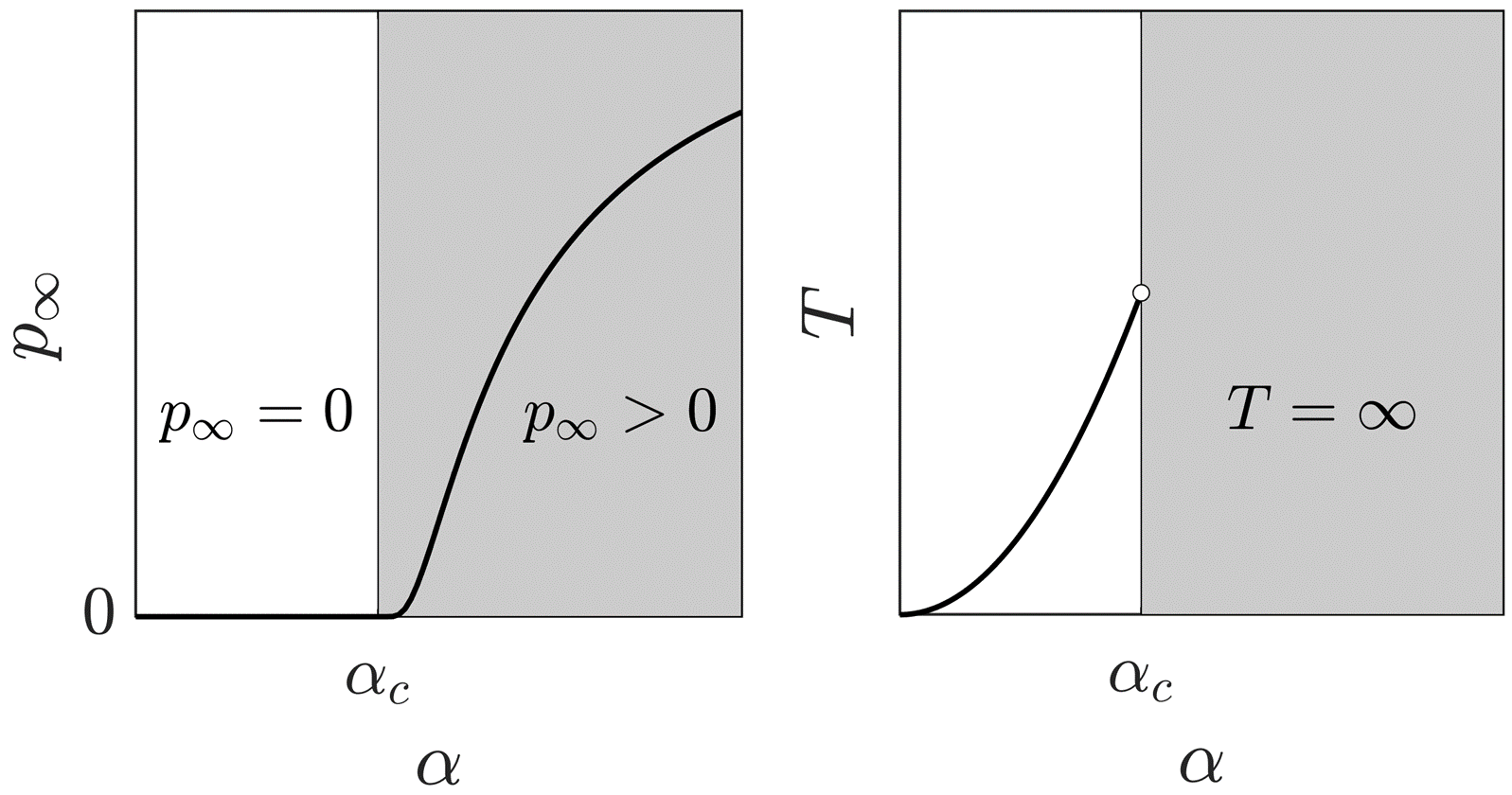}
		\caption{The transitions in $p_\infty$ and $T$ at $\alpha_c$. Here we illustrate the predicted transitions according to the theory. The upper plot shows that $p_{\infty}$, the likelihood to live forever, is $p_\infty=0$ when $\alpha<\alpha_c=1+1/\lceil S/2 \rceil$ and finite for $\alpha>\alpha_c$, see Eqs. (\ref{eq: pinf critic}) and (\ref{eq: alpha_c vs s}). At $\alpha_c$ there is an essential singularity in $p_\infty$, see Eq. (\ref{eq: exp critic}). The bottom plot shows the transition of the expected life time, $T$, at $\alpha_c$. The transition in $T$ is abrupt, where for   $\alpha<\alpha_c$ we find that $T$ always reaches a finite value and for $\alpha>\alpha_c$, we find $T\to\infty$, see Eq. (\ref{eq: T critic}).}	

		\label{fig: illustration of transitions}	

\end{figure}

As $\alpha \to \alpha_c^+$, it can be found that $C(\alpha) \sim \frac{1}{\alpha-\alpha_c}$, 
{ since converting the sum in Eq.~\eqref{eq:ln-pinf} to an integral gives  $C(\alpha)\approx -B-A \int_{D_0}^{\infty}x^{(1-\alpha)k}\mathrm{d} x \sim A D_0^{(\alpha-\alpha_c)k} / \left( (\alpha_c-\alpha)k \right)$ }.
Thus we can obtain the scaling relationship near criticality as
\begin{equation}
p_{\infty} \sim \exp \left(-\frac{b}{\alpha-\alpha_c}\right)
\label{eq: exp critic}
\end{equation}
where $b$ is some positive constant and $p_{\infty} \to 0$.
This implies that  $p_\infty$ undergoes a continuous transition  at $\alpha_c$ (see Fig. \ref{fig: illustration of transitions}), and at $\alpha_c$ there is an essential singularity. 

\noindent

To better understand the nature of the transition at $\alpha_c$, we consider $\nn$, the number of distinct sites the forager visits (equal to the number of meals the forager eats). We can define the likelihood of the forager consuming $\mathcal{N}$ meals i.e., visiting $\mathcal{N}$ distinct sites, as 
\begin{equation}
p_{\mathcal{N}} = (1-\phi_{\mathcal{N}+1})\prod_{D=1}^{\mathcal{N}}\phi_D
\label{eq:phi-n}
\end{equation}
with $p_{\mathcal{N}}=p_{\infty}$ for $\mathcal{N}\to\infty$. 
We can substitute Eq.~\eqref{eq: FL} into Eq.~\eqref{eq:phi-n}, and follow the same steps that were used to reach Eq.~\eqref{eq:ln-pinf} and \eqref{eq: exp critic} to obtain for large $\mathcal{N}$ the scaling
\begin{equation}
 p_{\mathcal{N}} \sim \mathcal{N}^{(1-\alpha )k} 	 e^{-A/((1-\alpha)k+1) \mathcal{N}^{(1-\alpha)k+1}}.
 \label{eq:p_N}
\end{equation}
Because of the exponential decay we find
\begin{equation}
\nn = \sum_{\mathcal{N}=1}^{\infty} \mathcal{N} p_{\mathcal{N}} = 
\begin{cases}
\text{finite}, \quad  \alpha < \alpha_c\\
\infty, \hspace{0.8cm} \alpha > \alpha_c
\end{cases},
\label{eq:avg-N}
\end{equation}
where $\alpha_c$ is the same as for Eq.~\eqref{eq: alpha_c vs s}. This may seem trivial, however it is worth noting that even if $p_\infty\to0$, the \emph{average} number of steps could still potentially diverge, however our above reasoning demonstrates that this does not happen. 

We thus conclude that  $\nn$ has a transition at $\alpha_c=1+1/k=1 + \frac{1}{\lceil S/2 \rceil}$ where for $\alpha<\alpha_c$, $\nn$ has a finite value while for $\alpha>\alpha_c$, $\nn$ diverges. Next, we wish to determine if this transition is continuous or abrupt.
We show in the Appendix that when $\alpha \to \alpha_c^-$ then $\nn$ always approaches some finite value, which indicates that there is an abrupt transition in $\nn$ at $\alpha=\alpha_c$. \\

\noindent
It is easy to see that $T$ behaves similarly to $\nn$ because $\nn +1 \leq T \leq (\nn+1 )S$. Therefore 
\begin{equation}
T = 
\begin{cases}
\text{finite}, \quad  \alpha < \alpha_c\\
\infty, \hspace{0.8cm} \alpha > \alpha_c
\end{cases}.
\label{eq: T critic}
\end{equation}
We illustrate the nature of the transitions, and the relevant regimes as determined by the theory in Fig.~\ref{fig: illustration of transitions}. \\

\noindent
To summarize, our derivations demonstrate that there exists a value $\alpha_c$ at which phase transitions occur in  $p_{\infty},\nn, \text{ and } T$ and that this critical $\alpha_c$ is given by Eq.~\eqref{eq: alpha_c vs s}. The transition for $p_{\infty}$ is continuous, and the scaling behavior near criticality is exponential. In contrast, surprisingly, the transition for $\nn$ and $T$ is  discontinuous or abrupt.\\  

\section{Two dimensions}
\begin{figure}[htp]
	\includegraphics[width=0.49\textwidth]{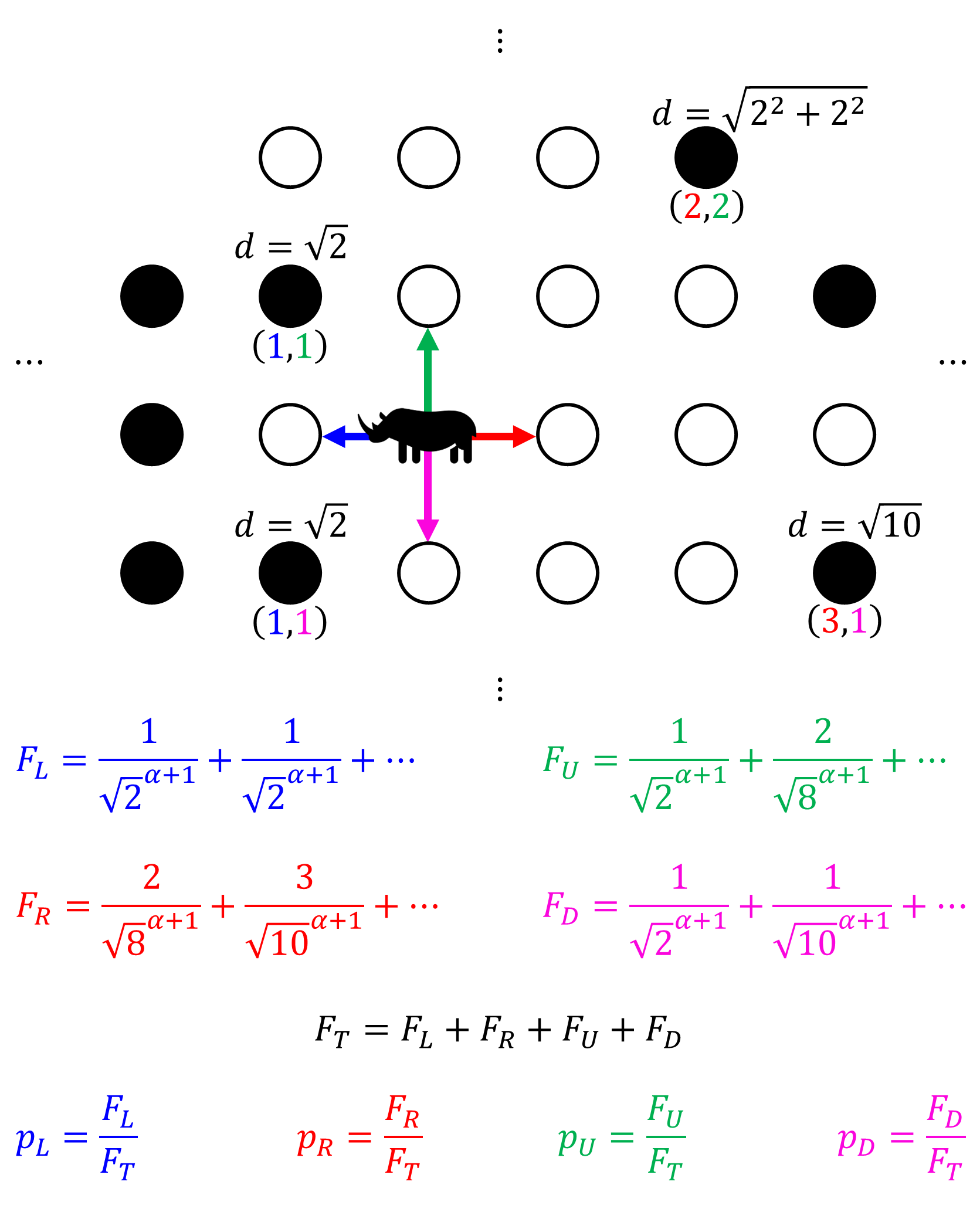}
	\caption{Model demonstration in 2D. In two dimensions there are four possible directions, $F_{L,R,U,D}$ and $d$ is found by $d=\sqrt{\Delta x^2+ \Delta y^2}$. The probability, $p_i$, to walk in each direction is found by dividing each direction by the total smell $F_T$.}
	\label{fig:model-2d}
\end{figure}

We will now consider the more ecologically relevant case of two dimensions. In this case, we must revise our definition for the smell due to a different food distribution. There are now 4 possible directions (on a square lattice) that the forager can move (rather than 2 directions for one dimension), and thus 4 different values of smell must be calculated. We assume that  food located at distance $(\Delta x,\Delta y)$ from the forager contributes $\frac{1}{\sqrt{(\Delta x^2+dy^2)}^{\alpha}}\left(\frac{\Delta x}{\sqrt{\Delta x^2+ \Delta y^2}},\frac{\Delta y}{\sqrt{\Delta x^2+\Delta y^2}}\right)$ to the smell. We sum all of the smells in the positive and negative directions of $x$ and $y$ and stochastically choose one of the 4 proportionally to their smell (see Fig.~\ref{fig:model-2d}).

\begin{figure}[htp]
	\centering
	\subfloat{%
		\includegraphics[width=0.49\textwidth]{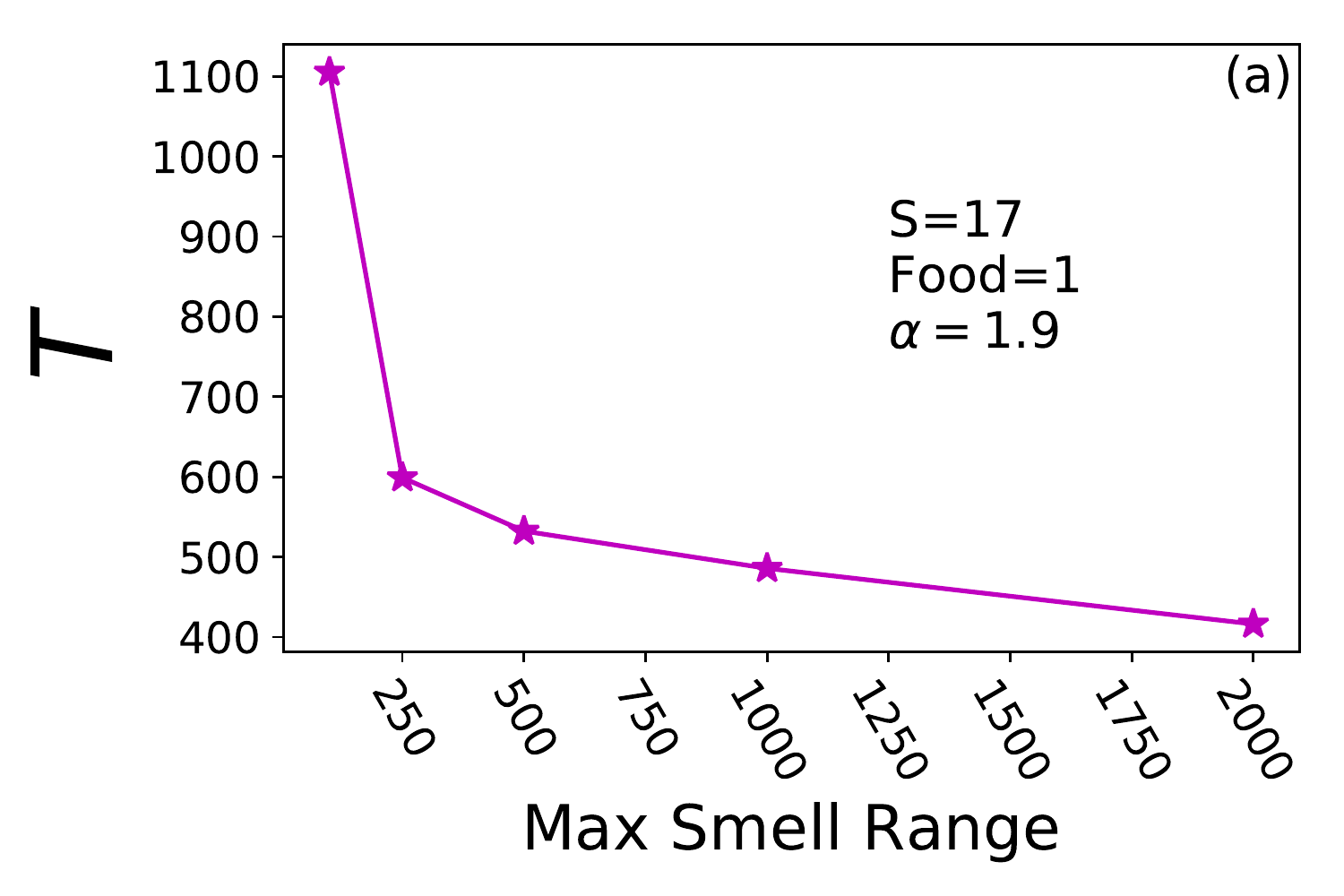}  		\label{fig:2d2}
	} \vfill
	\subfloat{%
		\includegraphics[width=0.49\textwidth]{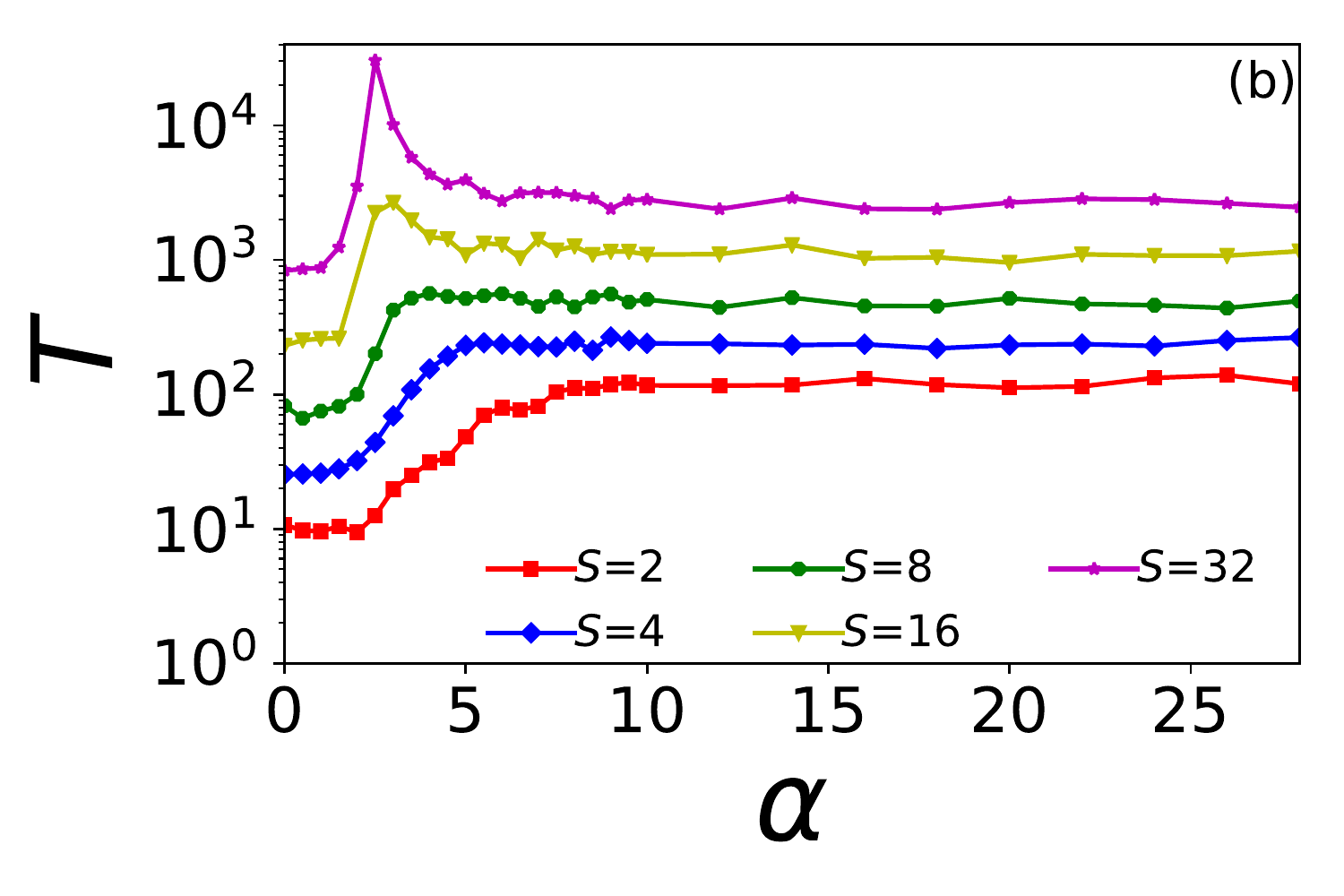}  		\label{fig:2d1}
	}

	\caption{Plots for a forager with long-range smell in two dimensions.  \textbf{(a)} We show that varying this maximum smell range actually significantly effects the forager's lifetime and thus one must carry out larger computations to assess the finite size issues. For smell range $\leq1000$ the overall system size is $10^3\times 10^3$, while for $2000$ the system size is $2\cdot10^3\times2\cdot 10^3$. In \textbf{(b)} we show the mean lifetime of the forager, $T$, as a function of $S$ for a two dimensional space of $10^3 \times 10^3$ sites with periodic boundary conditions. Our results suggest (at least for large $S$) that there exists a maximum in the forager lifetime. All points are averaged over $100$ realizations.}
	\label{fig:forager-2d}
\end{figure}

\subsection{Results}
To gain more intuition into the effect of smell, we convert to polar coordinates where the distance from an individual food item is $(r\cos\theta,r \sin\theta)$, where $r=\sqrt{x^2+y^2}$ and $\theta=\tan^{-1}\left(\frac{y}{x}\right)$. These coordinates allow us to simplify the total smell in a particular direction. For example, we can approximate the total upper smell, $u$, with the integral $\int_{0}^{\pi}\int_{r_0}^{\infty} 1/r^\alpha \sin \theta \, r \, \mathrm{d}r\,\mathrm{d}\theta=2\int_{r_0}^{\infty} r^{1-\alpha}\mathrm{d}r$. This integral diverges for $\alpha \leq 2$ and the same will be true for the lower, left, and right directions. Therefore, for $\alpha \leq 2$ our model converges to the model of the uniformly random walker forager \cite{benichou2014depletion} since the forager will experience infinite smell in all directions. As for the one-dimensional case, here too when $\alpha\to \infty$ the forager will behave like the completely greedy forager smelling only the nearest neighbors \cite{bhat2017does}.

Another point worth noting, is that our useful idea in one dimension where we recognized that the entire tail of the smell can be approximated as $\zeta(\alpha)$ will not work in two dimensions since the forager does not need to walk in an orderly manner, but rather can `snake' throughout a region. Therefore analytic results are difficult for 2D and in our simulations we actually construct a finite-size system and check which lattice sites have food in order to determine the smell in each direction. 
  
We begin by examining the finite-size effects on the lifetime of the forager for fixed $S$ and $\alpha$ near $\alpha=2$. Since in practice we cannot simulate an infinite system, we construct systems of size $l\times l$ with periodic boundary conditions. For $l\to \infty$, we expect that for all $\alpha\leq 2$, the forager lifetime should be constant and equal to that of the random walking forager \cite{benichou2014depletion}. Since the simulations are quite heavy, we consider setting some maximum distance up to which we will consider the impact of food, whereas food at greater distances will be ignored. In Fig. ~\ref{fig:2d2} we show that for $\alpha=1.9<2$, the forager lifetime decreases significantly as this maximal smell range increases. This can be understood by recognizing that the far away food has a significant effect on the total sum and causes the forager to have a negligible difference between the smell in all directions. When only nearby food is considered, the forager has a slight bias to go in a direction that overall has slightly more food leading to a longer lifetime, see Fig.~\ref{fig:2d2}. We choose 1000 as our linear-size limit, since it balances between being reasonably computationally feasible and giving results that are sufficiently close to those expected for the infinite limit.

In Fig.~\ref{fig:2d1} we plot the forager lifetime as a function of $\alpha$ for different values of $S$. We observe that from $\alpha=0$ until $\alpha\approx 2$ the forager lifetime is constant, as expected. This is since the smell in each direction is diverging and thus the walker behaves randomly. As $\alpha$ further increases, we see that the forager lifetime increases. For larger values of $S=16, 32$, the forager lifetime reaches a peak before dropping to a nearly constant value. For smaller $S$, the forager lifetime simply increases until it stabilizes at a constant value and it is not clear if a maximal value exits. This constant value for large $\alpha$ can be recognized as approaching the limit of total greed in  \cite{bhat2017does}. There, it was found that near the limit of total greed in 2 dimensions, the forager lifetime actually decreases as the forager becomes more greedy because the forager forms deserts and becomes `trapped' inside them.

The peak observed for $S=16, 32$ in Fig.~\ref{fig:2d1} is also related to a similar effect, since if $\alpha$ is too large, then the forager only considers nearby food, but if $\alpha$ is somewhat smaller, the forager will also consider farther away food and will thus preferentially avoid forming the desert and becoming trapped. For small $S$, we do not observe the peak, possibly due to its magnitude being smaller with our statistics unable to detect it.

\section{Discussion}
We have studied a forager who walks preferentially according to the overall smell of food in a given direction. We assume different values of the power-law exponent, $\alpha$,  governing the decaying of smell according to the distance of the food and find that this exponent significantly effects the forager lifetime. In one dimension above a certain critical $\alpha_c$, the forager lives for infinite time and almost always walks in the same direction. The value of $\alpha_c$ decreases with the time, $S$, that the forager can live without food. For a forager in two dimensions, we find evidence that there is an optimal value of $\alpha$ for which the forager lifetime is maximal. Furthermore, in the limits of sufficiently small $\alpha$ (in one dimension $\alpha\leq1$ and in two dimensions $\alpha\leq2$), our results recover those of \cite{benichou2014depletion}, whereas for $\alpha\to\infty$ our results recover those of a forager with total greed in \cite{bhat2017does}. 

Overall, our results provide intuition on how long-distance smell effects the lifetime of a forager seeking to optimize food consumption. Further work could also explore cases where originally only some fraction of sites contain food, consider sites with multiple portions of food, and consider how smell could influence prior results on the myopic forager who only eats if it is sufficiently close to death \cite{benichou2018optimally}.
%
%


\bibliographystyle{unsrtnat}
\bibliography{Forager}

\clearpage
\section{Appendix}
\setcounter{figure}{0}
\setcounter{equation}{0}
\renewcommand{\thefigure}{A.\arabic{figure}}
\renewcommand{\theequation}{A.\arabic{equation}}
\subsection{Verifying the behavior of $p_D(x)$ and $q_D(x)$}

\begin{figure}[htp]
	\centering	
	\subfloat{
		\includegraphics[width=0.43\textwidth]{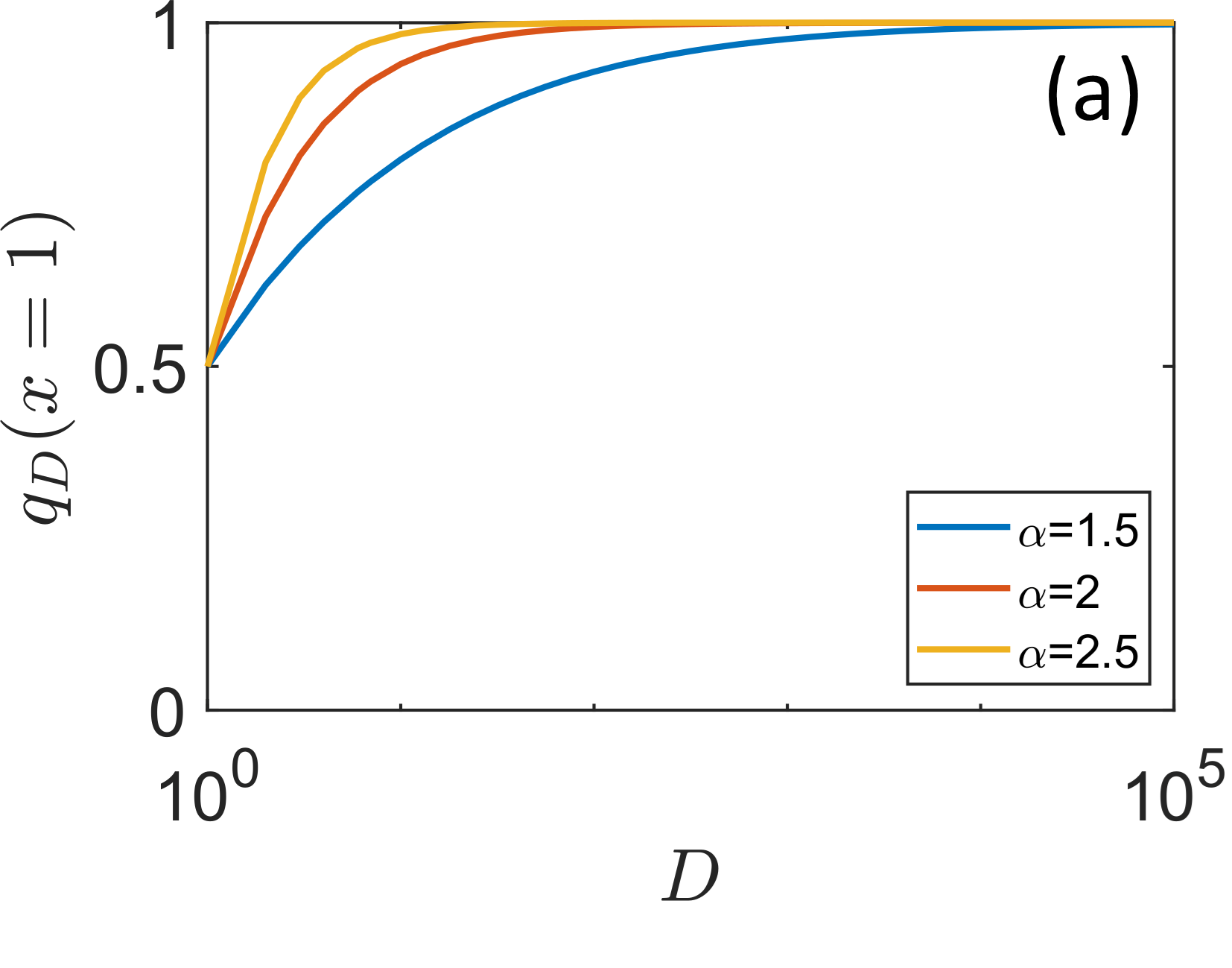}
		\label{fig: q vs D x1}
	}
	\hfill
	\subfloat{
		\includegraphics[width=0.43\textwidth]{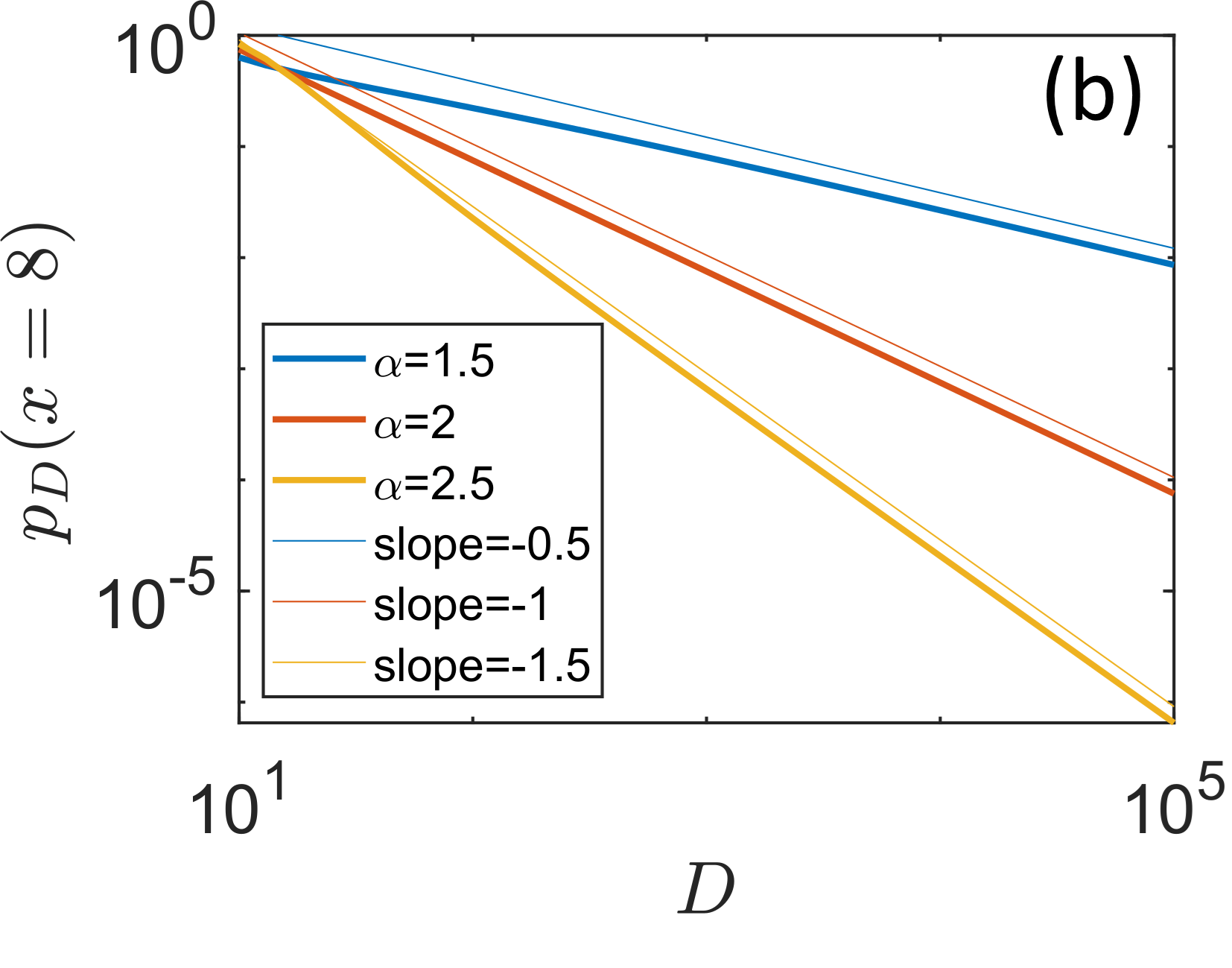}
		\label{fig: p vs D x8}		
	}
	\caption[short text]{
		\textbf{(a)} Here we show the likelihood of a forager with an infinite line of food  in one direction and $D$ empty food sites in the other direction (after which all sites have food), to choose to walk in the direction of the infinite line of food. In \textbf{(b)}, we show the same as Fig.~\ref{fig: p vs D x1} from the main text but for $x=8$. As there, the dotted lines represent slopes with the values given in the legend. The scaling properties confirm the analytic result of Eq.~\eqref{eq:pL}..
	}
	\label{fig:p-L-sup}
\end{figure}
Here we further verify our results from Eq.~\eqref{eq:pL}. We show how $q_D(x=1)$ approaches a likelihood of $1$ as the size of the desert increases. Similarly, we verify Fig.~\ref{fig: p vs D x1} from the main text also for $x=8$ and find that the scaling properties are still the same for large $D$.

\subsection{Behavior of $\nn$ in 1D near criticality}
Here we study what happens to the mean number of consumed food units, $\nn$ at $\alpha=\alpha_c$ and $\alpha \to \alpha_c^-$. 
\noindent At $\alpha=\alpha_c=1+1/k$, Eq.~\eqref{eq:p_N} becomes 
\begin{equation}
p_{\mathcal{N}} \approx A\mathcal{N}^{-1} e^{-B+A \ln D_0 -A \ln \mathcal{N}} \sim \mathcal{N}^{-A-1},
\label{Eq:p_N2}
\end{equation}
where $A$ and $B$ are the same constants as in Eq.~\eqref{eq:ln-pinf}.
We see from Eq.~\eqref{Eq:p_N2}, that $\nn$ converges if and only if $A>1$.  To find $A$ we can recognize from Eq. (\ref{eq: FL}) that
\begin{equation}
1-\phi_D \sim p_D^k q_D^{S-k} \sim A D^{(1-\alpha)k}.
\end{equation} 
The constant $A$ is the product  of the coefficient of $p_D$ in Eq.~\eqref{eq:pL}  and $n_k^S$, the number of starving paths with $k$ steps to the right.  
We will refer to the first part of this product, namely the coefficient of $p_D$ in Eq.~\eqref{eq:pL},  as $A_1$ and thus 
\begin{align*}
p_D(x) &\sim F_R/F_L\\
&\sim D^{1-\alpha} /[(\alpha-1)F_L(x)]\\
&\equiv A_1(x) D^{1-\alpha},
\end{align*}
which implies
\begin{equation}
A_1(x) = [(\alpha-1)F_L(x)]^{-1}.
\end{equation}
At criticality, we substitute $\alpha_c=1+1/k$ and get
\begin{equation}
A_1(x) = k/F_L(x).
\end{equation}
We will now calculate the second factor in $A$, namely $n_k^S$. 
$n_k^S$ is the number of paths from $x=1$ with $k$ of $S$ steps to the right which \textit{do not reach} $x=0$. This is equal to the \textit{total} number of paths from $x=1$ with $k$ of $S$ steps to the right \textit{minus} the number of paths from $x=1$ with $k$ of $S$ steps to the right that \textit{do reach} $x=0$. The paths that reach $x=0$ have a one-to-one correspondence with paths from $x=-1$ to the same end point, meaning $k+1$ steps to the right. Thus $n_k^S=\binom{S}{k}-\binom{S}{k+1}$ and $n_1^1=1$. \\

\noindent
Since $A_1(x)$ varies with $x$ we cannot find $A$, but we can provide bounds, which will be sufficient for our needs. These  are
\begin{equation}
A_1(1)^{k} n_k^S \leq A \leq A_1(k)^{k} n_k^S,
\end{equation}
We can now find $A$ at $\alpha=\alpha_c$. \\
For $S=1,2$ $k=1$ so $A=1/F_L(1)=1/\zeta(\alpha_c)=1/\zeta(2)\approx 0.61<1$, hence $\nn=\infty$. 
When $S=3,4$ then $k=2$ and we have $A\geq A_1(1)^{2} n_2^S = (2/F_L(1))^2\cdot2  = 8/\zeta(1.5)^2 \approx 1.2 >1$, hence $\nn<\infty$.
More generally, we find $A \geq A_1(1)^{k} n_k^S = k^k/(\zeta(1+1/k))^k \left( \binom{S}{k}-\binom{S}{k+1} \right)$ and for all $S>2$ we find that $A>1$, hence $\nn$ is finite.
For $S>2$, we can recognize that if at $\alpha_c$, $\nn$ is finite, then for $\alpha<\alpha_c$, $\nn$ is bounded because $\nn$ monotonically increases with $\alpha$ . 

Since $A<1$ for $S=1,2$ we still must determine for $\alpha \to\alpha_c^-$, if $\nn$ is bounded or if it diverges. For $S=1,2$, we know $k=1$ i.e., a single step away from the food will lead to the death of the forager. We now consider $\nn$ for $\alpha=\alpha_c-\epsilon$, where we will take the limit of small $\epsilon$. In this regime we can convert  Eqs.~\eqref{eq:p_N}-\eqref{eq:avg-N} to an integral as
\begin{equation}
\nn \approx C_1 +C_2 \int_{\mathcal{N}_0}^{\infty} \mathcal{N}^{\epsilon }\exp \left(-\frac{A}{\epsilon}\mathcal{N}^{\epsilon }\right)\mathrm{d}\mathcal{N}.
\end{equation}
where $C_1$ and $C_2$ are some positive constants, $A$ is the same constant as before, and $\mathcal{N}_0$ is sufficiently large such that Eq.~\eqref{eq:p_N} is satisfied.
By the transformation $\left\{ u=\frac{A}{\epsilon }\mathcal{N}^{\epsilon } \right\}$  
we obtain 
\begin{equation}
\nn -C_1 \sim \Gamma(z+1,Az) / (Az)^{z}
\label{eq:n_gamma_func}
\end{equation} where $\Gamma$ is the upper incomplete gamma function and $z=1/\epsilon$. We now want to evaluate $\nn$ in the case $\epsilon \to 0$ which implies $z\to \infty$. 
Using the above result $A = 0.61$, we find numerically that $\nn-C_1\to0$ in Eq.~\eqref{eq:n_gamma_func}  for $z\to \infty$. Thus, we conclude that $\nn$ approaches a finite number for $\alpha \to \alpha_c^-$ for all $S$ (including $S>2$ due to our earlier arguments) and therefore $\nn$ is bounded for $\alpha<\alpha_c$. This implies a discontinuous transition in $\nn$ at $\alpha_c$, since as $\alpha\to\alpha_c$ we do not find that there is any asymptotic divergence of $\nn$. 

\end{document}